\documentclass[submission, Proceedings]{SciPost}

\hypersetup{
    colorlinks,
    linkcolor={red!50!black},
    citecolor={blue!50!black},
    urlcolor={blue!80!black}
}

\usepackage{tikz}
\usetikzlibrary{arrows,backgrounds,snakes}
\usetikzlibrary{decorations.markings}
\usetikzlibrary{decorations.pathmorphing,decorations.markings}
\usepackage[compat=1.1.0]{tikz-feynman}
\usetikzlibrary{shadows}

\usepackage[bitstream-charter]{mathdesign}
\DeclareSymbolFont{usualmathcal}{OMS}{cmsy}{m}{n}
\DeclareSymbolFontAlphabet{\mathcal}{usualmathcal}

\def\dataspace{(0,0) circle [x radius=2, y radius=1.7]}
\def\dataset{(0.3,-0.5) circle [x radius=1.3, y radius=0.5]}
\def\modelpar{(7,0) circle [x radius=1.8, y radius=0.45]}
\def\invdata{(6.75,0.2) circle [x radius=1.8, y radius=0.82]}
\def\modelspace{(6.7,-0.2) circle [x radius=2.5, y radius=2.2]}

\colorlet{circle edge}{blue!50}
\colorlet{circle area}{blue!20}

\tikzset{filled/.style={fill=circle area, draw=circle edge, thick},
    outline/.style={draw=circle edge, thick}}

\newcommand{\ndata}{N_\mathrm{dat}}

\newcommand{\levtwo}{z}

\newcommand{\model}{g[\theta]}

\newcommand{\repind}{(k)}
\newcommand{\cov}{C}
\newcommand{\invcov}[1]{\cov^{-1}_{#1}}

\newcommand{\repchis}{{\chi^2}^{\repind}}

\newcommand{\diffreptwo}{\left( \model^{\repind} - \levtwo^{\repind} \right)}

\begin{document}

\begin{center}{\Large \textbf{
An Overview of Lattice Results for Parton Distribution Functions
}}\end{center}

\begin{center}
Luigi Del Debbio\textsuperscript{1,*}
\end{center}

\begin{center}
{\bf 1} 
Higgs Centre for Theoretical Physics,\\
School of Physics and Astronomy, \\
The University of Edinburgh, \\
Edinburgh EH9 3FD, Scotland \\
* luigi.del.debbio@ed.ac.uk
\end{center}

\begin{center}
\today
\end{center}


\definecolor{palegray}{gray}{0.95}
\begin{center}
\colorbox{palegray}{
  \begin{tabular}{rr}
  \begin{minipage}{0.1\textwidth}
    \includegraphics[width=23mm]{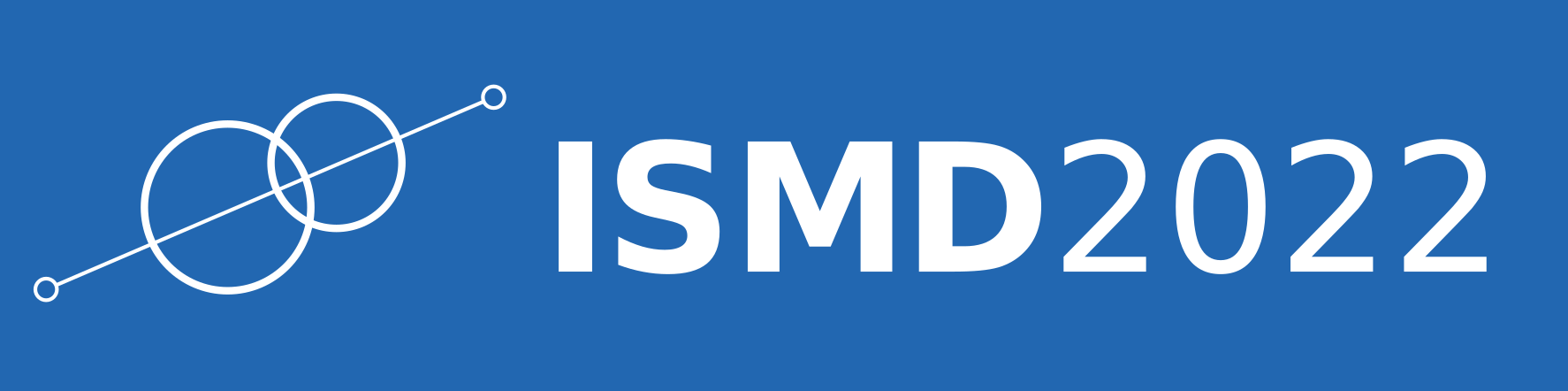}
  \end{minipage}
  &
  \begin{minipage}{0.8\textwidth}
    \begin{center}
    {\it 51st International Symposium on Multiparticle Dynamics (ISMD2022)}\\ 
    {\it Pitlochry, Scottish Highlands, 1-5 August 2022} \\
    \doi{10.21468/SciPostPhysProc.?}\\
    \end{center}
  \end{minipage}
\end{tabular}
}
\end{center}

\section*{Abstract}
{\bf
  Following a ground-breaking proposal by Ji~\cite{PhysRevLett.110.262002}, numerical
  simulations of Quantum Chromo Dynamics (QCD) on a Euclidean lattice
  have provided new, valuable information on the structure of
  hadrons. In this talk, we briefly review the lattice approach to
  the reconstruction of parton densities, highlighting the connection 
  between lattice observables
  and parton densities, with a focus on theoretical
  issues. Since parton distributions are extracted from
  lattice data by solving an inverse problem, we discuss some of the
  difficulties that affect these determinations and how they can be
  formulated in a Bayesian framework.  }

\vspace{10pt}
\noindent\rule{\textwidth}{1pt}
\tableofcontents\thispagestyle{fancy}
\noindent\rule{\textwidth}{1pt}
\vspace{10pt}

\section{Introduction}
\label{sec:intro}
A detailed understanding of the structure of hadrons, and of nucleons
in particular, is central to the current and future experimental
explorations of nature at particle accelerators. Two notable examples
are the forthcoming runs of the Large Hadron Collider (LHC) at CERN
and the experimental programme at the Electron Ion Collider (EIC),
which is under construction in the US. As far as the LHC is concerned,
the focus for future runs will be on increasing the luminosity and
therefore the precision of the experimental measurements. Our current
understanding of proton collisions being based on factorization
theorems, a detailed knowledge of Parton Distribution Functions (PDFs)
is necessary in order to be able to exploit the increase in
statistical precision. The EIC, on the other hand, is designed
specifically to investigate the internal structure of nucleons by
probing the dynamics of quarks and gluons in head-on collisions of
protons (or heavier nuclei) with a beam of electrons. The interplay 
between theory and experiment will ultimately provide the most accurate
description of hadronic matter.  

As a result, we expect in the near future significant progress in the
determination of PDFs and in our understanding of the strong force in
general, leading to a description of nucleons and possibly nuclei and
atoms from first principles, i.e. from the QCD lagrangian. Lattice
QCD is by now a mature tool for accurate QCD predictions and will play
an active role in this area. 

In this proceedings, we start by reviewing the field-theoretical
definition of parton distributions, explaining how the nucleon
structure is encoded in field correlators that are defined on the
light-cone and can be computed in quantum field theories. We will then
introduce a simple toy model in order to highlight the relation
between these light-cone quantities and Euclidean correlators that are
amenable to Monte Carlo simulations. We will see that the Euclidean
correlators, after being properly renormalized, can be treated exactly
like experimental data for which there exists a factorization
theorem. 

As a consequence of these factorization theorems, PDFs can only be
extracted by solving an inverse problem. In the final part of this
review, we discuss some features of inverse problems and their
solution in a Bayesian framework.

\section{Theoretical Background}

\begin{figure}[h]
  \centering
  \includegraphics[scale=0.4]{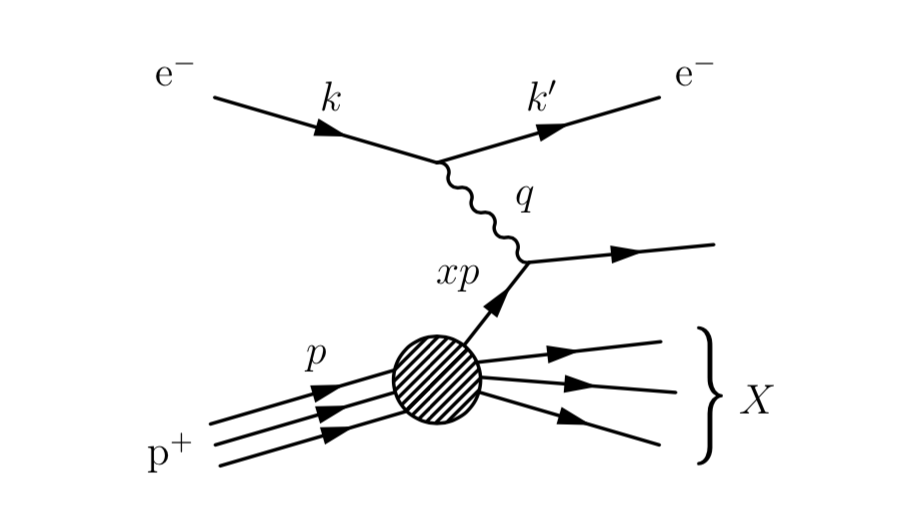}  
  \caption{DIS process, where a lepton with momentum $k$ scatters of a
    nucleon with momentum $p$. In the deep inelastic regime, the four-
    momentum squared $Q^2=-q^2$ is much larger than the typical
    hadronic scale $\Lambda^2$.}
  \label{fig:DISPlot}
\end{figure}

The current understanding of inelastic scattering processes that
involve hadrons is based on factorization theorems. In kinematical
regimes where there is a separation of scales, factorization theorems
allow us to separate {\em soft}, non perturbative contributions from
high-energy partonic dynamics, up to corrections given by powers of
the ratio of the scales of the problem. The typical example of such
processes is Deep Inelastic Scattering (DIS), where a lepton scatters
off a nucleon through the exchange of a photon with a very large
transferred momentum, as shown in Fig.~\ref{fig:DISPlot}. 

The hadronic contribution to the cross section is encoded in the
tensor
\begin{align*}
    H_{\mu\nu} &= \sum_X (2\pi)^D \delta\left(p-p_X-q\right)\,
                 \langle p | J_\mu(0) | X\rangle
                 \langle X | J_\nu(0) | p\rangle \, ,
\end{align*}
which contains the matrix element of the electromagnetic current
between the nucleon in the initial state $|p\rangle$ and a generic
hadronic state $|X\rangle$ in the final state. Using covariance under
Lorentz transformations, the non perturbative dynamics due to the
strong interactions is encoded in two form factors, 
 \begin{align*}
    H_{\mu\nu}
    =& F_1(x,Q^2) \left(\frac{q_\mu q_\nu}{q^2} - g_{\mu\nu}\right) +
       F_2(x,Q^2) \left(p + \frac{1}{2x} q\right)_\mu
       \left(p + \frac{1}{2x} q\right)_\nu\, .
  \end{align*}

\begin{figure}[h]
  \centering
   \includegraphics[scale=0.5]{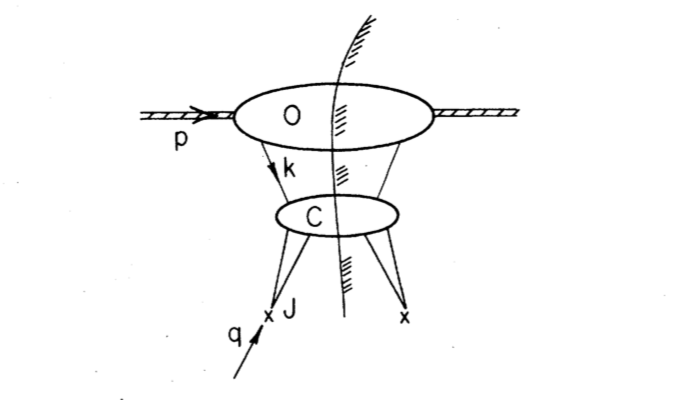}  
   \caption{Leading contribution to the hadronic tensor in the deep
     inelastic limit. Hard momenta flow in the lower part of the
     diagram, which yields the perturbative coefficient function
     $C_i$, while the upper part contains all the soft lines. The
     leading contribution is characterized by having only two
     fermionic lines connecting the upper and the lower part of the
     diagram. Diagrams with more lines between these two subdiagrams are
     suppressed by powers of $Q^2$. Diagram from
     Ref.~\cite{Collins:1980ui}.}
  \label{fig:LeadingDIS}
\end{figure}

  In the {\em deep inelastic} regime, the norm of the transferred
  four-momentum is large compared to the typical hadronic scale,
  $\Lambda^2/Q^2 \ll 1,$ and the form factors can be factorized as
  \begin{align}
    \label{eq:FactThmF}
    F_i(x,Q^2) = \int_x^1 \frac{d\xi}{\xi}\, C_{i}(\xi,Q^2,\mu^2)
    f_R(x/\xi,\mu^2) + \mathcal{O}(\Lambda^2/Q^2)\, ,
  \end{align}
  where the coefficient function $C_i$ is related to a partonic cross
  section that can be computed in perturbation theory, while the
  effect of the strong interactions in the nucleon is encoded in the
  PDF $f_R$. The suffix $R$ in this expression indicates that the PDFs
  are physical, finite quantities, defined in a particular
  renormalization scheme. Corrections to the factorized expression
  are suppressed by powers of $\Lambda^2/Q^2$. The core of the
  argument that underlies factorization is the fact that the hadronic
  matrix elements are dominated by diagrams as the one shown in
  Fig.~\ref{fig:LeadingDIS}. See e.g. 
  Ref.~\cite{Collins:1980ui} for a concise discussion in the context 
  of a scalar field theory. 

Looking at the structure of the diagram, it is clear that the non
perturbative contribution to the hadronic tensor is given by the
matrix element of a fermion bilinear (corresponding to the two fermion
lines that connect the two halves of the diagram) between nucleon
states. When all the details are worked out, the PDF is obtained as 
the integral of the matrix element of the fermion bilinear along the 
light-cone direction $z^-$,
\begin{align}
  \label{eq:PDFMatEl}
  f(x) = \int \frac{dz^-}{2\pi}\, e^{i (xp^+) z^-}\, \langle p | \bar
  \psi(-z^-/2)\, \Gamma \lambda^A\, \mathcal{U}\, \psi(z^-/2) | p
  \rangle\, ,
\end{align}
where $\Gamma$ determines the spin structure of the bilinear, while
$\lambda^A$ is a matrix in flavor space and $\mathcal{U}$ is a Wilson
line from $-z^-/2$ to $z^-/2$ which guarantees gauge invariance. The
expression in Eq.~(\ref{eq:PDFMatEl}) yields the {\em bare} PDF. It is 
UV-divergent and needs to be renormalized in the desired scheme in order 
to match the function $f_R(x)$ that appears in the factorization formula,
Eq.~(\ref{eq:FactThmF}). 

Similar arguments apply for less inclusive quantities, which lead to
the introduction of Wigner functions, the basis for the definition of
Generalized Parton Distribution Functions (GPDFs) and Transverse
Momentum Dependent Parton Distribution Functions (TMDs),
\begin{align*}
  W(x,k_\perp,b_\perp) 
  &= 
    \int \frac{d^2\Delta_\perp}{(2\pi)^2} e^{-i \Delta_\perp \cdot b_\perp}\,   
    \frac12 \int \frac{dz^- d^2z_\perp}{(2\pi)^3} 
    e^{i \left[(xp^+) z^- - k_\perp \cdot z_\perp\right]}\, \times \\
  & \quad \quad \times \langle p + \frac{\Delta_\perp}{2} | 
    \bar \psi(-z/2)\, \Gamma \lambda_A\, \mathcal{U}\, \psi(z/2) |
    p - \frac{\Delta_\perp}{2}
  \rangle \, .
\end{align*}
We see explicitly from this equation that Wigner functions are defined as 
off-diagonal elements of field bilinears, with the possibility of 
injecting a transverse momentum $k_\perp$.

\section{Lattice QCD and PDFs: a toy model}
\label{sec:another}

At first sight, it is difficult to imagine that lattice QCD, defined
in Euclidean space, can provide information on the PDFs, since
the latter are
extracted from correlators evaluated along the light-cone. Indeed, for
several decades, the only quantities that were amenable to Monte Carlo
simulations were the moments of the PDFs, which are defined from the
OPE as matrix elements of local operators, see
e.g. Ref.~\cite{PhysRevD.4.1059,PhysRevD.9.416}. The seminal work in
Ref.~\cite{PhysRevLett.110.262002} introduced Euclidean matrix
elements, where the fields in the fermion bilinear are separated in a
purely spatial direction. These matrix elements allow the extraction
of the PDFs using a factorization formula. In this respect, they are
on the same footing as physical observables, and indeed they can be
incorporated in phenomenological fits, as discussed in
Refs.~\cite{Cichy:2019ebf,DelDebbio:2020rgv}. It is pedagogical to
study these Euclidean correlators in a simple scalar field theory,
along the lines of the work originally presented in
Ref.~\cite{Collins:1980ui}. In this simple framework, the conceptual
steps can be understood without being distracted by complex
calculations. Hence, we are going to consider the matrix elements 
\begin{align}
  \label{eq:ScalarEuclCorr}
  \mathcal{M} = \langle p | \phi(z) \phi(0) | p\rangle\, ,
\end{align}
in a scalar $\phi^3$ field theory in $D$ dimensions, 
\begin{align}
  S[\phi] = 
  \int d^Dx\, \left(
  \frac12 \partial_\mu \phi(x) \partial^\mu \phi - 
  \frac12 m^2 \phi(x)^2 - 
  \frac{\lambda}{3!} \phi(x)^3
  \right)\, .
\end{align}
In order to establish the necessary factorization theorem, we need to
\begin{enumerate}
\item Renormalize the bilinear for $z$ along the light-cone and for a
  Euclidean separation. 
\item Find a relation between these two renormalized quantities, in the form
  of a factorization theorem, with IR-finite coefficient functions.  
\end{enumerate}
The results summarised here were originally discussed in
Ref.~\cite{DelDebbio:2020cbz}.

\subsection{Renormalization}

\begin{figure}[h]
  \centering
    \includegraphics[scale=0.5]{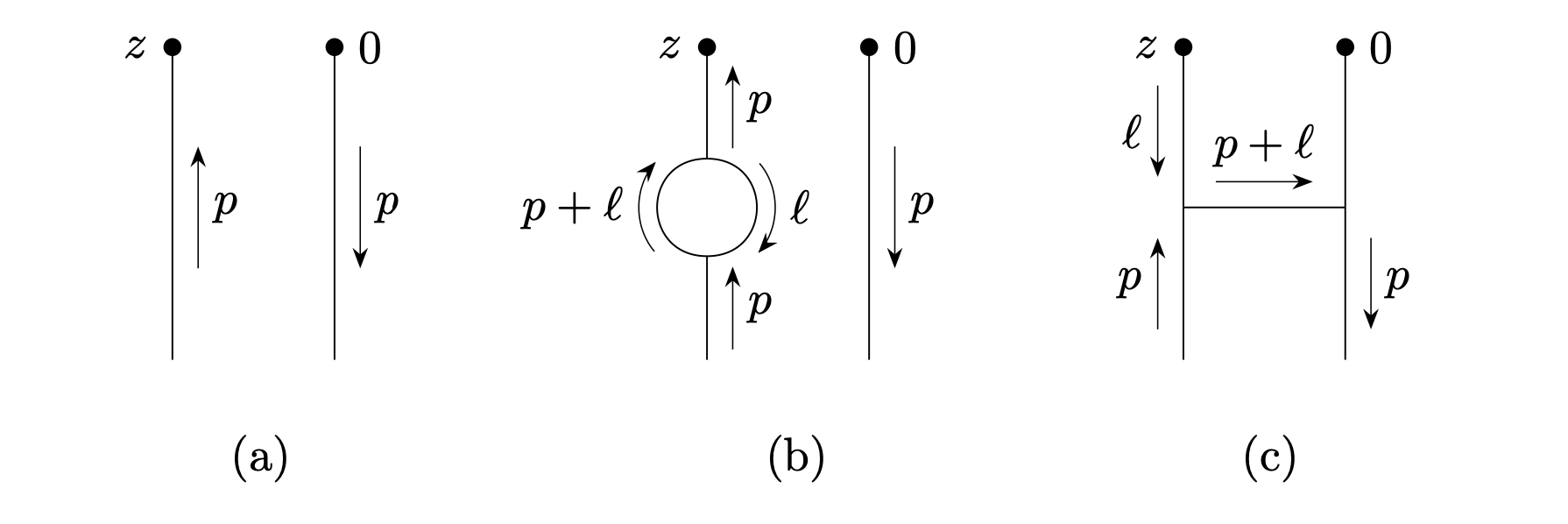}    
  \caption{Tree level and one-loop diagrams needed for the
    renormalization of the field bilinear $\phi(z)\phi(0)$.}
  \label{fig:MatElemOneLoop}
\end{figure}

In order to address the renormalization of the bilinears, we are going
to consider their matrix elements between partonic states
(i.e. elementary fields $\phi$), which we denote
$\hat{\mathcal{M}}(\nu,z^2)$. Lorentz invariance implies that the
matrix element can only depend on the scalar quantities $\nu=p\cdot z$
and $z^2$. We use dimensional regularization and
$\overline{\mathrm{MS}}$ for renormalization.  The tree level and
one-loop diagrams that are needed to renormalize the field bilinear
are summarised in Fig.~\ref{fig:MatElemOneLoop}. 

At tree-level, as shown in diagram (a), the matrix element only
depends on $\nu$,
\begin{align*}
    & \widehat{\mathcal{M}}_a(\nu, z^2) = \exp[-i p\cdot z] = \exp [-i \nu
    ] = \widehat{\mathcal{M}}^{(0)}(\nu, 0)\, . 
\end{align*}
Note that there is no difference between the light-cone and the Euclidean
expression at tree level. In both cases the result only depends on the variable $\nu$.

For a light-cone separation, $z^2=0$, the bilinear at one-loop is
logarithmically divergent. The renormalized matrix element is
\begin{align}
  \label{eq:LCRenorm}
  \widehat{\mathcal{M}}_R\left(\nu; \mu^2\right)
  &= \left[1+ \frac{\alpha}{6} \left(\log\frac{m^2}{\mu^2} +
    b\right) \right]
    \widehat{\mathcal{M}}^{(0)}\left(\nu, 0\right) \\ 
  &+ \alpha\,  \int_0^1 dx\,\left(1-x\right)\,
    \log\frac{\mu^2}{m^2\left(1-x+x^2\right)}
    \widehat{\mathcal{M}}^{(0)}\left(x\nu, 0\right)\, ,
\end{align}
where the one-loop corrections are proportional to the coupling
$\alpha=\lambda^2/(64\pi^3)$.  The multiplicative renormalization in
the first line is the usual renormalization of the field, which
appears in diagram (b). The convolution in the second line comes from
diagram (c) in Fig.~\ref{fig:MatElemOneLoop} and is the typical
divergence that appears when such bilinears are concerned.  Note that 
we have suppressed the dependence on $z^2$ since in the light-cone 
case we always have $z^2=0$. The renormalized light-cone matrix element
is the quantity that is directly related to the PDF. 

For a Euclidean separation, $z^2=-z_3^2$, the finite value of $z^2$
acts as a regulator and the one-loop calculation is finite after the 
usual renormalization of the field $\phi$,
\begin{align}
  \label{eq:EuclRenorm}
    \widehat{\mathcal{M}}_R\left(\nu, z_3^2; \mu^2\right)
    &= \left[1+ \frac{\alpha}{6} \left(\log\frac{m^2}{\mu^2} +
      b\right) \right]
      \widehat{\mathcal{M}}^{(0)}\left(\nu, 0\right) \\ 
    &~+\alpha \int_0^1 dx\,\left(1-x\right)\,
      2K_0\left(m z_3\right)
       \widehat{\mathcal{M}}^{(0)}\left(x \nu, 0\right)\, ,
\end{align}
where $K_0$ is a Bessel function, which is singular only at the origin $z_3=0$.

\subsection{Factorization}

In order to obtain a factorization theorem, we expand the expression
for the Euclidean correlator for $m z_3 \ll 1$, 
\begin{align*}
  2K_0\left(M z_3\right) =
  - \log\left(m^2z_3^2\right) + 2 \log\left(2e^{-\gamma_E}\right) +
  \mathcal{O}\left(m^2 z_3^2\right)\, .
\end{align*}
We see explicitly that the function in the convolution is
IR-divergent when $m\to 0$. However, this divergence matches exactly
the divergence in the convolution for the light-cone quantity. When
relating the two renormalized expressions, the dependence on $m^2$
cancels and we obtain a relation that is IR-finite, as required in a
factorization theorem. By inverting the relation between the light-cone
renormalized correlator and the renormalized PDF $f_r$, we can 
write 
\begin{align}
  \label{eq:FactLatCorr}
    &\widehat{\mathcal{M}}_R\left(\nu, -z_3^2; \, \mu^2\right) =
      \int_{-1}^{1} d\xi \, \tilde{C}\left(\xi\nu,\mu^2 z_3^2 \right)
      \hat{f}_R\left(\xi,\mu^2\right) + O(m^2 z_3^2) \\
    &\tilde{C}\left(\xi\nu,\mu^2 z_3^2 \right) =
      e^{i\xi\nu} - \alpha\int_0^1 dx \, \left(1-x\right)
      \log\left( \mu^2 z_3^2\frac{e^{2\gamma_E}}{4} \right) e^{i x
      \xi\nu}\, .
\end{align}
The left-hand side of this equation is a Euclidean correlator, which can be
computed using standard techniques in Monte Carlo simulations of QCD. The
right-hand side is a convolution that involves the PDF as defined from the
light-cone correlator. It is interesting to notice that the dependence on $z^2$
only appears at $O(\alpha)$. This is the equivalent of the well-known violations
of Bjorken scaling in DIS. Note also that the factorization formula receives
corrections that are powers of $m z_3$, i.e. factorization works at small
distances, with power suppressed corrections, just like in DIS, with small
distances replacing large momentum transfer. 

There are a multitude of lattice observables that have been defined starting
from the Euclidean correlator in position space discussed here, see e.g.
Ref.~\cite{Cichy:2021ewm,Scapellato:2021uke} for recent reviews and exhaustive
references. All of these lattice observables are related to light-cone
correlators and therefore to PDFs following the steps outlined above. First the
lattice quantities need to be properly renormalized and then the renormalized
observable -- extrapolated to the continuum limit -- is matched to PDFs via some
factorization theorem. After this procedure has been completed the lattice
observables are exactly on the same footing as experimental observables, e.g.
like the structure functions $F_i$ introduced above in the context of 
DIS~\cite{DelDebbio:2020cbz,Monahan:2018euv}. 

\section{Bayesian Inverse Problems}
\label{sec:BayesInv}

\begin{figure}[h]
  \centering
   \begin{tikzpicture}
    \draw[outline] \dataspace node [xshift=-20, yshift=30] 
      {$\mathcal Z =\mathbb{R}^N$};
    \draw[outline] \dataset node [xshift=-10] {$A$}; 
    \draw[outline, rotate around={45:(7,0)}] \modelpar 
      node [xshift=20, yshift=25] {$\mathcal{F}$};
    \draw[outline, rotate around={155:(6.75,0.2)}] \invdata 
      node [xshift=-27, yshift=22] {$\mathcal{G}^{-1} A$}; 
    \draw[outline] \modelspace 
      node [xshift=20, yshift=-40] {$\mathcal M(\mathcal X, \mathcal Y)$};
    \node[anchor=south] at (0,2) {data space}; 
    \node[anchor=south] at (6.75,2) {model space}; 
    \draw [<-, red] plot [smooth] coordinates {(0.5,-0.5) (3.5,0.8) (7,0)} 
      node [xshift=-105, yshift=30] {$z=\mathcal G(f)$};
    \node [below] 
      at (0.5,-0.5) {$z$};
    \node [below] at (7,0) {$f$};
  \end{tikzpicture}    
  \caption{Schematic representation of an inverse problem. 
  On the left of the figure we see the space of data, with some dataset $A$. 
  On the right we have the space of measurable functions 
  $\mathcal M(\mathcal X, \mathcal Y)$ and the subset $\mathcal F$ 
  that we explore with a given parametrization.}
  \label{fig:InverseScheme}
\end{figure}
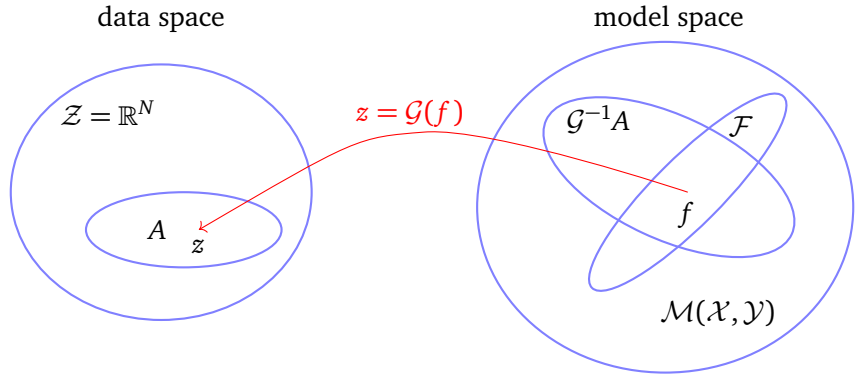

Determining the PDF from a finite set of lattice data using
Eq.~(\ref{eq:FactLatCorr}) is a typical example of an inverse problem. Lattice
simulations can provide a finite number of input points, which are used to
constrain a function, i.e. an element of an infinite-dimensional space. Such
problem is clearly ill-defined and the solution does depend on a number of
assumptions that need to be made. The formulation of the problem is summarised
as a sketch in the diagram in Fig.~\ref{fig:InverseScheme}. We aim to determine
a function $f$, in this particular case a function of one real variable $x$. The
model space, denoted $\mathcal M$ in the figure, is an infinite-dimensional
space. The data set $A$ is a dicrete set, each point in the data set is some
functional of the function $f$, $z=\mathcal G(f)$. Clearly the system is
underdetermined, since it would require an infinite amount of data in order to
uniquely determine the function $f$. A more constructive way to phrase the
problem can be formulated in a Bayesian framework~\cite{DelDebbio:2021whr}. 
Our knowledge, or prejudice, about the function $f$ is encoded in some prior 
$p(f)$, which is then updated using the dataset:
\begin{equation}
  \label{eq:BayesInverseEq}
  p(f|A) \propto p(A|f) p(f)\, ,
\end{equation}
where on the right-hand side we have introduced the likelihood function
$p(A|f)$. In principle one could try to work in infinite-dimensional spaces and
define a probability measure. While this is a possibility, it is easier to
reduce the problem by parametrizing the function $f$ and work with a probability
density in the finite-dimensional space of the parameters that define the model.
Clearly the choice of the parametrization induces a bias that needs to be 
taken into account. 

\subsection{NNPDF fits of lattice data}
  \label{sec:NNPDFLatFit}

As explained in the previous section, lattice data are on the same footing as
experimental input into PDFs fits. In a series of papers, some of the lattice
data have been incorporated in the general fitting framework developed by the
NNPDF collaboration~\cite{Cichy:2019ebf,DelDebbio:2020rgv}. It is worthwhile to
emphasise that the lattice data are handled like any other dataset in NNPDF,
with no need to adjust the methodology. The only input needed is a robust
estimate of the statistical covariance and the systematic errors. This is
modeled by considering the data, $z$, as stochastic variables distributed
according to a multi-dimensional Gaussian distribution, centred at the value 
of the experimental measurement $Z$, with a
covariance $C$, which we denote as
\begin{equation}
  \label{eq:YGauss}
  z \sim \mathcal N(Z,C)\, .
\end{equation}

\paragraph{Parametrization} In the NNPDF formalism, the parametrization of the
function $f$ is provided by neural networks, see Ref.~\cite{NNPDF:2021njg} for
the details of the latest implementation. A sufficiently large architecture
provides a parametrization that is flexible enough to minize the functional bias. 
We denote the neural net parametrization
as $g[\theta]$, where $\theta$ is the set of parameters (biases and weights of
the neural network). 

\paragraph{Posterior Distribution.} The posterior distribution in the space of
functions -- i.e. in the space of functions that are parametrized by the neural
networks -- is described by a Monte Carlo set of replicas. The replicas
implement a bootstrap propagation of the statistical fluctuations of data into
the space of functions~\cite{EfroTibs93}. Each replica $z^{(k)}$ is obtained by
generating a set of pseudo-data,
\begin{equation}
  \label{eq:DataGeneration}
  z^{(k)} = Z + \varepsilon^{(k)},~~ k=1,\ldots N_{\mathrm{rep}}\, ,
\end{equation}
where $\varepsilon^{(k)}$ are distributed according to $\mathcal N(0,C)$.

The set of replicas yields an ensemble of pseudo-data that reproduces the
statistical distribution of the experimental data as encoded in the covariance
matrix. For each replica, we perform a fit, i.e. we minimize a loss function,
subject to constraints~\footnote{The details of the training procedure are
very important, but beyond the scope of these proceedings. We advise the 
interested reader to scrutinise the algorithm with great care, and 
refer them to the detailed explanations in Ref.~\cite{NNPDF:2021njg}.}
\begin{align}
    \repchis[\theta] &= \frac{1}{\ndata} \sum_{ij} \diffreptwo_i \invcov{ij} \diffreptwo_j + \mathrm{priors}\, , \\
    \theta^{(k)} &= \arg\min_{\theta} \repchis[\theta]\, . 
\end{align}
Hence for each replica, we obtain a parametrization of the PDFs. This set of
fitted parameters, $\left\{\theta^{(k)}; k=1, \ldots, N_{\mathrm{rep}}\right\}$,
yields the desired posterior distribution in the space of functions. Note that
this ensemble of replicas is an example of importance sampling of the posterior
distribution in the space of functions parametrized by the neural networks. 

\begin{figure}[h]
  \centering
  \minipage{0.45\textwidth}
    \includegraphics[width=6cm]{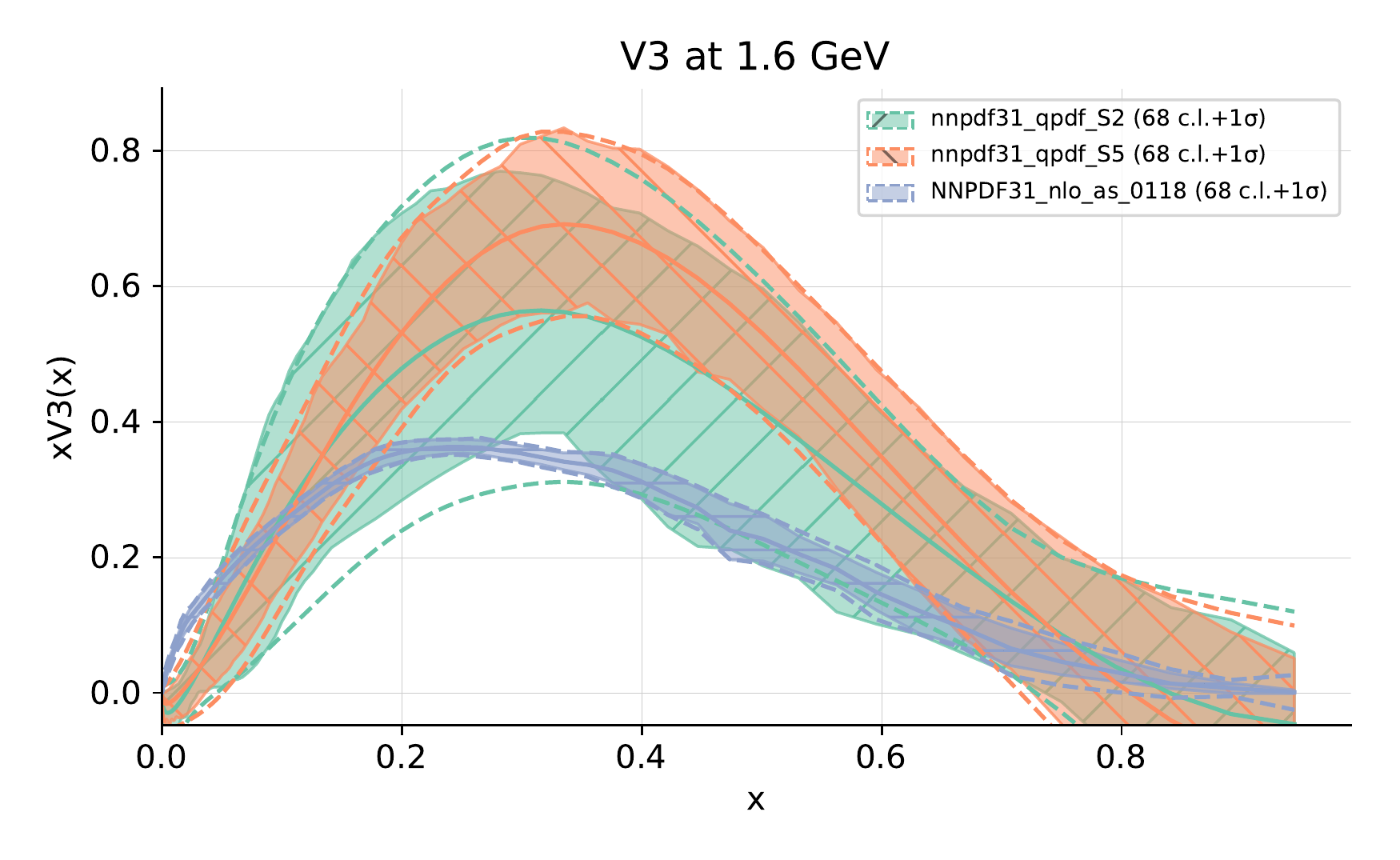}  
  \endminipage 
  \hspace{.05\textwidth}
  \minipage{0.45\textwidth}
    \includegraphics[width=6cm]{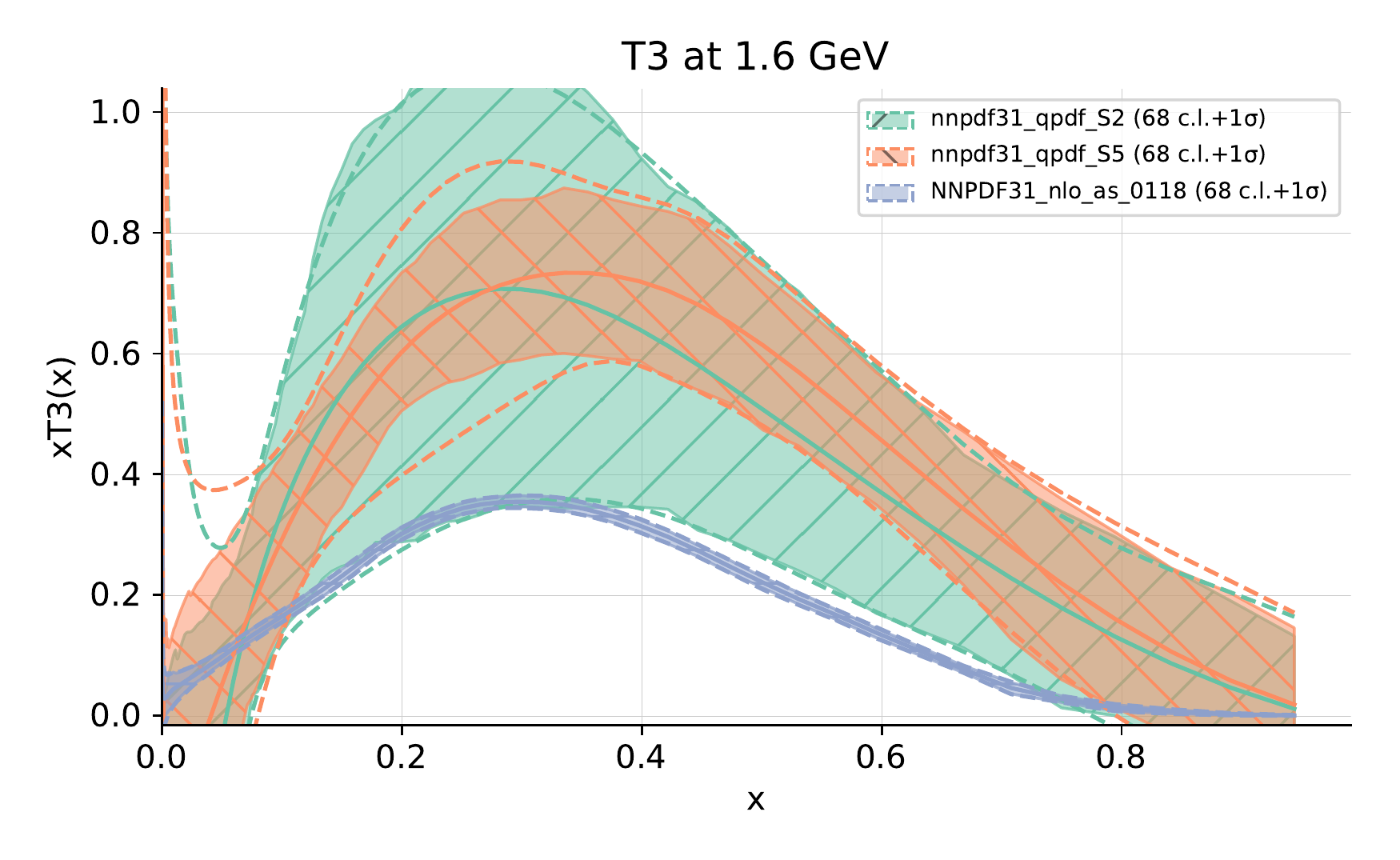}  
  \endminipage\hfill 
  \caption{PDFs extracted from lattice data using the NNPDF
    methodology. Results from Ref.~\cite{Cichy:2019ebf}. The NNPDF fit to
    experimental data is shown in blue for comparison. }
  \label{fig:NNPDFFitResOne}
\end{figure}

\begin{figure}[h]
  \centering
  \includegraphics[scale=0.4]{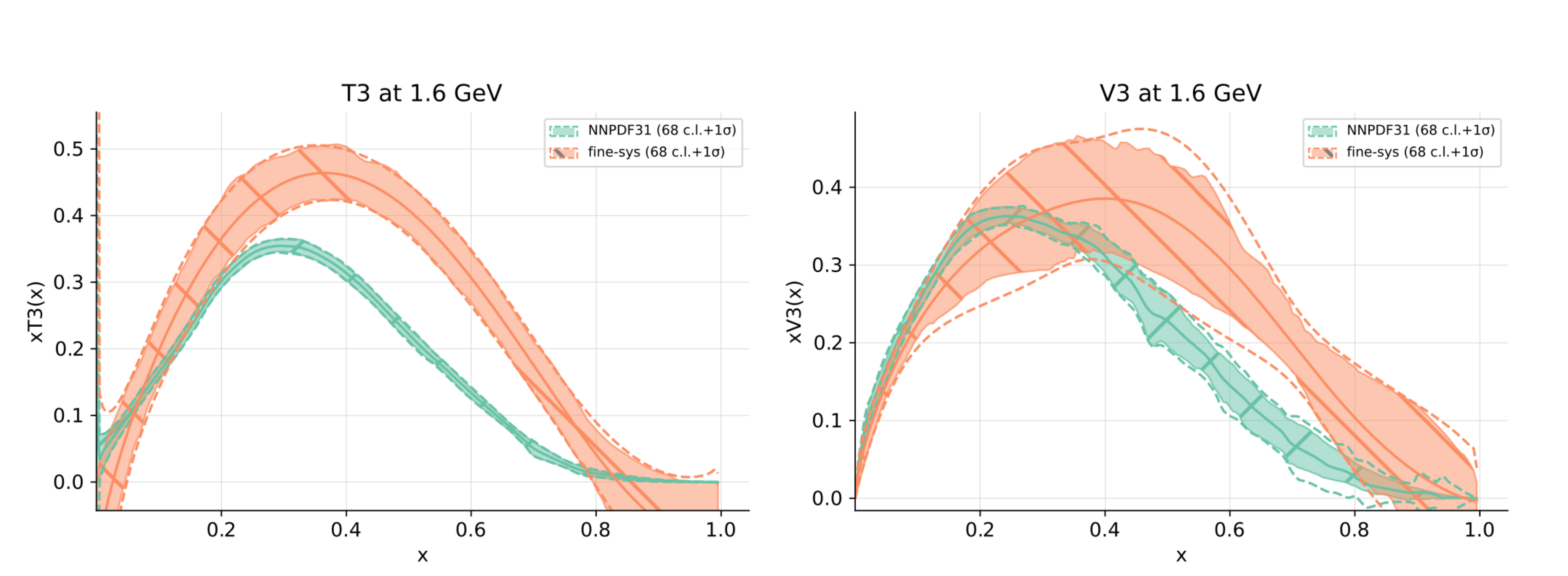}
  \caption{PDFs extracted from lattice data using the NNPDF methodology. 
  Results from Ref.~\cite{DelDebbio:2020cbz}. The NNPDF fit to experimental data is shown in 
  green in this plot.}
  \label{fig:NNPDFFitResTwo}
\end{figure}

Results of the fits to lattice data were presented in
Refs.~\cite{Cichy:2019ebf,DelDebbio:2020cbz}. A few examples are reported in
Fig.~\ref{fig:NNPDFFitResOne} and~\ref{fig:NNPDFFitResTwo}. It is important to
remark that the lattice error in these analysis is dominated by systematic errors.

Despite the large errors, it is remarkable that a fairly small number of lattice
data points provides a powerful constraint for the PDFs. It is clear that
improved lattice simulations, with better control of systematic errors, could
target the kinematical regions were the current uncertainties are large and have
a significant impact. 

\section{Recent fits}
  \label{sec:RecentFits}

Since the analyses in Refs.~\cite{Cichy:2019ebf,DelDebbio:2020cbz}, a large
number of new simulations have appeared that have measured a broader variety 
of observables, and have improved both the statistical and
systematic errors. These more recent simulations have also focused on helicity
distributions, TMDs and GDPs, which are less precisely determined from fits to
experimental data. These are areas where the lattice simulations are most likely
to have a lasting impact. We do not have time to make an exhaustive review of
contemporary results, for which we refer the reader to the recent proceedings of
the Lattice conference~\cite{Scapellato:2021uke}. In order to entice the 
audience, we report here a few results from recent publications. This is not an 
exhaustive list of result, but does give an idea of the potential for lattice 
QCD. 

\begin{figure}[h]
  \centering
  \includegraphics[scale=0.3]{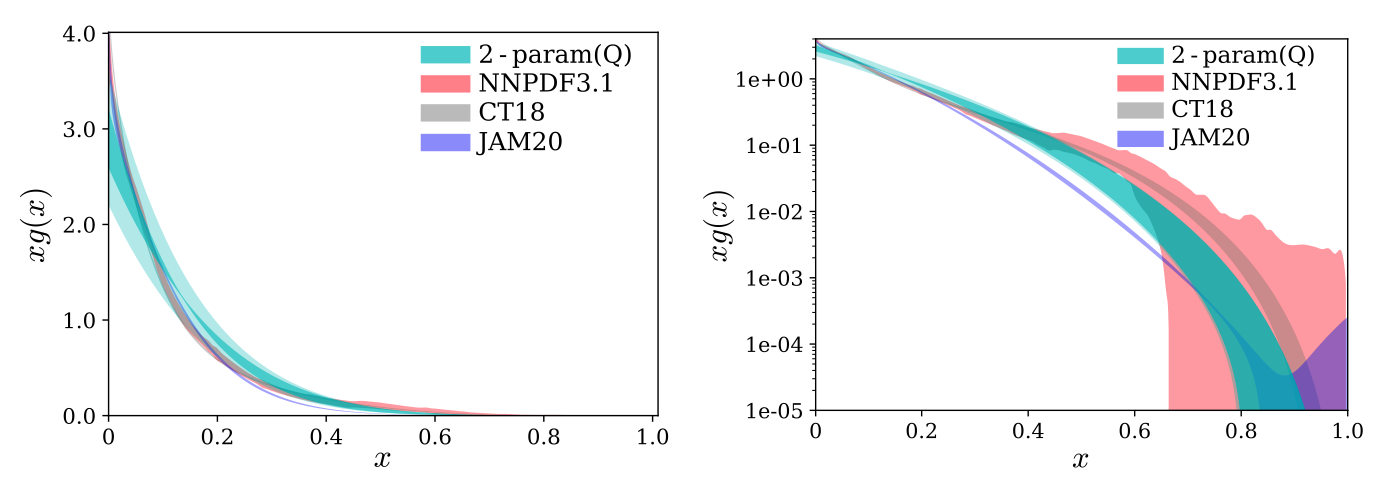}
  \caption{Unpolarized gluon PDF from Ref.~\cite{Khan_2021}. Lattice results 
  are compared with the results of fitting experimental data.}
  \label{fig:HadStrucGluon}
\end{figure}

\paragraph{Unpolarised Gluon.} The HadStruc collaboration has performed a
careful study of the unpolarised gluon distribution~\cite{Khan_2021}. The study
is interesting in many respects. In particular, the study takes into account the
mixing of the gluon and the singlet bilinears under renormalization and
introduces a new parametrization of the PDFs for their fits. The result is
reported here in Fig.~\ref{fig:HadStrucGluon}. Note that the pion mass in this
work is still relatively large, $m_\pi\simeq 358~\mathrm{MeV}$.

\begin{figure}[h]
  \centering
  \includegraphics[scale=0.45]{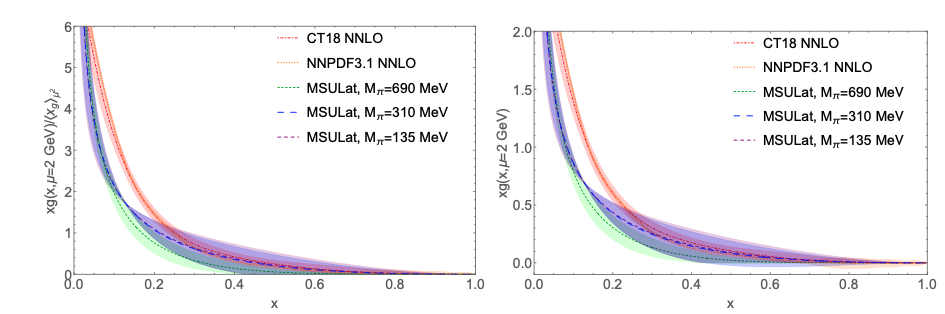}
  \caption{Unpolarized gluon PDF from Ref.~\cite{Fan_2021}. Lattice results 
  are compared with the results of fitting experimental data.}
  \label{fig:MSULatGluon}
\end{figure}

\paragraph{Another Unpolarized Gluon.} An independent determination of the
unpolarised gluon distribution is presented in Ref.~\cite{Fan_2021}. Despite the
fact that a wider range of pion masses is explored, there are still systematic
discrepancies between these determinations and the fits to experimental data,
which deserve to be clarified. It would be interesting to perform a combined
analysis of these results using the NNPDF methodology as explained in the
previous Section. These determinations are the benchmarks that the lattice
results need to satisfy in order to show that systematic errors are under
control. Results are shown in Fig.~\ref{fig:MSULatGluon}.

\begin{figure}[h]
  \centering
  \includegraphics[scale=0.35]{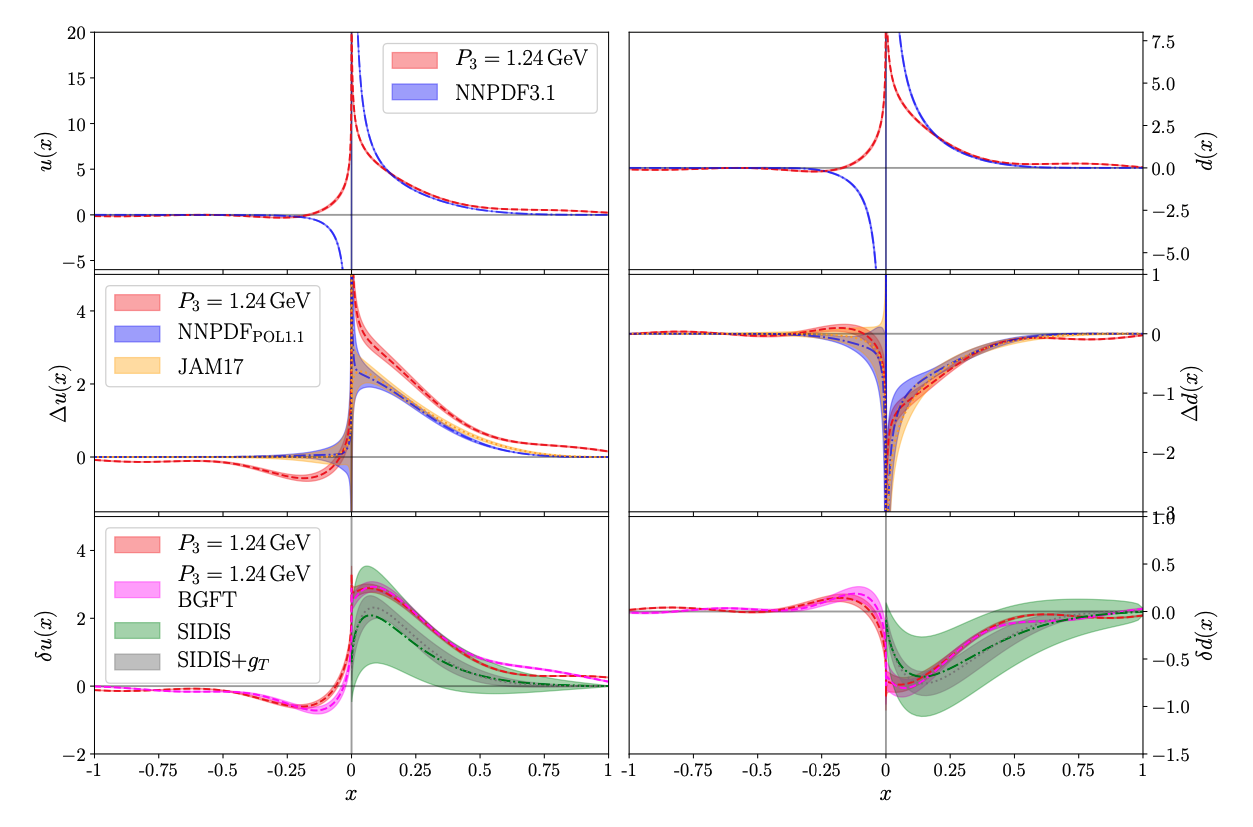}
  \caption{Helicity and transversity distributions from 
  Ref.~\cite{Alexandrou:2022ign}.
  Lattice results are compared with the results of fitting experimental data.}
  \label{fig:ETMCHelAndTra}
\end{figure}

\paragraph{Helicity and Transversity.} Ref.~\cite{Alexandrou:2022ign} is one
example of a study that goes beyond the unpolarized distributions and addresses
the determination of helicity and transversity parton distributions. These
distributions are less constrained by experimental data, so that the impact of
lattice simulations here is likely to be highly significant. Results are
reported in Fig.~\ref{fig:ETMCHelAndTra}.

\section{Conclusion}
  \label{sec:Conclusion}

Since the original work in Ref.~\cite{PhysRevLett.110.262002}, there has been a
momentous activity in the lattice community in order to extract PDFs from
numerical simulations of lattice QCD. Following a series of well-founded 
criticisms in Refs.~\cite{Rossi:2017muf,Rossi:2018zkn} the theoretical status of these studies
has been carefully examined, and it is now clear that lattice data in Euclidean spacetime
can be related to light-cone PDFs by factorization theorems after the lattice
observables have been properly renormalized and extrapolated to the continuum
limit. 

In this respect, lattice data can be treated exactly like any other experimental
data. The extraction of the PDFs requires the solution of an ill-defined inverse
problem, whose result depends on the choice of priors. This is a delicate point,
which needs to be analysed with great care. A Bayesian formulation facilitates
the explicit formulation of the assumptions underlying the prior disribution.
Having clarified the prior, the posterior distribution is given by Bayes
theorem, and can be sampled e.g. by Monte Carlo methods as suggested in the
NNPDF approach. 

As the high-energy physics community is driven towards precision analyses at
hadronic colliders, a faithful determination of the errors on the Parton
Distribution Functions becomes increasingly important. We expect to see
significant progress in this area, especially from the synergy of different
approaches. 
  
\section*{Acknowledgements}
LDD is grateful to the collaborators who were involved in the studies used for
these proceedings. T~Giani and C~Monahan contributed significantly to shaping
the author's understanding of the theoretical issues. 

\paragraph{Funding information}
LDD is supported by the U.K. Science and Technology Facility Council (STFC)
grant ST/P000630/1.



\bibliography{ldd_ismd22.bib}


\end{document}